\documentclass[english,a4paper,11pt]{article}
\usepackage[margin=3cm]{geometry}
\usepackage[latin1]{inputenc}
\usepackage{babel}
\usepackage{bm}
\usepackage{amsmath,amsthm}
\usepackage{latexsym}
\usepackage{booktabs}
\usepackage[final]{graphicx}
\DeclareGraphicsExtensions{.jpg,.jpeg,.pdf,.png,.mps}
\usepackage{epsfig}
\usepackage[round]{natbib}
\setcitestyle{numbers}
\setcitestyle{square}

\usepackage{rotating}
\usepackage{color}
\usepackage{xcolor}
\usepackage{colortbl}
\usepackage[bitstream-charter]{mathdesign}
\usepackage[T1]{fontenc}
\usepackage{threeparttable}
\usepackage{lipsum}
\usepackage{url} 
\usepackage{hyperref}
\usepackage{lmodern}

\usepackage{subcaption}
\usepackage{comment}
\date{}
\hypersetup{hidelinks}

\usepackage{acronym}

\acrodef{RL}{Reinforcement Learning}
\acrodef{MDP}{Markov Decision Process}
\acrodef{MBDP}{Markov Birth-Death Process}
\acrodef{MMPP}{Markov-Modulated Poisson Process}
\acrodefplural{MMPP}[MMPPs]{Markov-Modulated Poisson Processes}
\acrodef{PMF}{Probability Mass Function}
\acrodef{MSS}{Micromobility Sharing System}
\acrodefplural{MSS}[MSS's]{Micromobility Sharing Systems}
\acrodef{MaaS}{Mobility as a Service}
\acrodef{ML}{Machine Learning}

\title{Regulating Spatial Fairness in a Tripartite Micromobility Sharing System via Reinforcement Learning}

\author{$\mathrm{Matteo \ Cederle}^\mathrm{1},  
	\  \mathrm{Marco \ Fabris}^\mathrm{1},
	\  \mathrm{Gian \ Antonio \ Susto}^\mathrm{1}$\\  
	$^\mathrm{1}$\small{\emph{University of Padova}}
}

\begin{document}
	\maketitle

\begin{abstract}
	
\noindent In the growing field of Shared Micromobility Systems, which holds great potential for shaping urban transportation, fairness-oriented approaches remain largely unexplored. This work addresses such a gap by investigating the balance between performance optimization and algorithmic fairness in Shared Micromobility Services using Reinforcement Learning.
Our methodology achieves equitable outcomes, measured by the Gini index, across central, peripheral, and remote station categories. By strategically rebalancing vehicle distribution, it maximizes operator performance while upholding fairness principles.~The efficacy of our approach is validated through a case study using synthetic~data.
\hspace{1cm}
\end{abstract}

\section{Introduction}\label{sec:intro}

Over the past decade, \acp{MSS} have become a vital part of urban transit, providing last-mile services and reducing pollution in terms of CO$_2$ emissions.
Rebalancing, which involves redistributing vehicles to meet demand within \ac{MSS} management, remains a major cost for operators~\cite{chiariotti2020bike}. 
In particular, a key challenge in \acp{MSS} is equity, as bikes and scooters are often concentrated in wealthier areas, leaving poorer neighborhoods underserved~\cite{smith2015exploring}.
While dockless systems have reduced some disparities~\cite{brown2021docked}, issues in vehicle distribution persist, with significant differences in availability across neighborhoods~\cite{mooney2019freedom}. 
This equity issue relates to spatial fairness, which seeks uniform resource allocation but often conflicts with demand-based optimization~\cite{soja2009city}.
For these reasons, we analyze the trade-off in dockless \acp{MSS} using a tunable \ac{RL} scheme that incorporates spatial fairness. 
The main contributions of this article are summarized below:  
\begin{itemize}  
	\item We propose a fairness-aware \ac{MSS} simulator by clustering areas into three categories, as detailed in Section \ref{sec:sys}; 
	\item To the best of our knowledge, this is the first study addressing fairness in \ac{MSS} operation and rebalancing rather than just planning (see Section \ref{sec:rl_appr});
    \item
    In Section \ref{sec:exp}, we demonstrate via Monte Carlo simulations an inherent trade-off between system performance and fairness levels achieved using a parametric family of RL strategies.
\end{itemize}  

\section{Background and modeling of dockless MSS's}\label{sec:sys}

A dockless \ac{MSS} is defined as a fully connected graph $\mathcal{G} = (\mathcal{V},\mathcal{E})$, where the set of nodes $\mathcal{V}$ represents a partition of the city map, with each node describing a relatively small area over which the number of vehicles is counted. $\mathcal{E} = \mathcal{V} \times \mathcal{V}$ instead denotes the set of connections between each pair of areas.
Moreover, it is crucial to observe that accurately modeling and predicting the dynamics of such networks in their entirety is not a computationally tractable problem for large \ac{MSS} services.
We then focus on a stochastic model of an individual service area, considering independent \acp{MMPP} for the arrivals and departures, which is consistent with experimental results on large sharing systems~\cite{chiariotti2018dynamic}.
The vehicle occupancy of each area then follows a left-censored continuous-time Markov Birth-Death Process (MBDP)~\cite{andronov2011markov}.

We will consider a system with $V=3$ service areas, which we divide in three categories according to common spatial patterns in European and US cities as listed in the following.
\emph{Central} areas in the city center, where large businesses and attractions are clustered, typically have a high traffic volume, with more arrivals than departures in the morning and the opposite in the evening due to commuter traffic.
\emph{Peripheral} areas, which are typically residential areas close to the center, have a lower density, and thus less traffic, but present an inverted pattern with respect to daily activities, i.e., more departures in the morning and arrivals in the evening. 
\emph{Remote} areas are typically underserved suburbs and lower-income communities. The traffic patterns are similar to peripheral areas, but with an even lower traffic volume due to the factors we highlighted above.

\subsection{Fairness metrics in MSS's}\label{ssec:fairness}

Most of the literature about equity in \acp{MSS} considers fairness for planning purposes and on a \emph{system-level perspective}~\cite{duran2020fair}:
the objective of this work is to consider fairness from the perspective of a single user, applying this metric for system rebalancing operations. 

A clear trade-off with rebalancing efficiency arises: areas with a lower traffic demand will still require rebalancing trucks to travel a certain distance, reducing the expected profit for the system with respect to more central, high-demand areas.
We thus consider the Gini index as a general fairness metric,
to assess the equality of access to services within a population. Its values span from 0 to 1, where 0 indicates a perfectly fair system and 1 indicates high unfairness. In our context, it is defined~as $g(x) = \frac{1}{2n^2\bar{x}} \sum\nolimits_{j=1}^{n} \sum\nolimits_{k=1}^{n} |x_j - x_k|,$
where $n$ is the number of categories, $x_k$ is the probability of service failure at finding an available vehicle in a given category, and $\bar{x}$ denotes the mean of quantities $x_k$ over $k=~1,\ldots,n$.

\section{Fairness-Oriented Reinforcement Learning Approach}\label{sec:rl_appr}

Our statistical model relies on an independence assumption: the \acp{MMPP} representing arrivals and departures in each area are assumed to be independent both from each other and from the processes in other areas. Naturally, this assumption is not verified in real systems, as trips usually begin in an area and end in another a few minutes later, but the approximation error is surprisingly low in large-scale systems~\cite{chiariotti2020bike}: any individual area makes up such a small fraction of the total traffic that local events have negligible effects elsewhere.
This independence property allows to consider individual rebalancing actions in different areas as separate problems, modeling the system as a transition and reward independent 
multi-agent Markov Decision Process:
actions from one agent have no effect on the state transitions of others.
The overall state of the multi-agent problem can be then factored into individual state components for each area.
The elements that constitute the state of each individual agent are the time of the day, i.e. morning or evening, the area type, i.e. \textit{central, peripheral or remote}, and the number of vehicles currently available in the area.
The action space for each agent is designed to
offer meaningful choices without overwhelming the agent with too many options. Actions include adding or removing up to 30 vehicles, by increments of~5.

The reward function $R(t)$ then models the objective of rebalancing operations, i.e., the system operator's profits and operational costs associated with the management of the \ac{MSS}. This economic interest is the combination of various factors. Firstly, the most significant cost in managing \acp{MSS} is represented by rebalancing itself: whenever a truck is dispatched to an area, the operator incurs a cost that is proportional to the centrality of the area. In order to consider the costs of rebalancing different areas and fairness issues between neighborhoods, we partition $\mathcal{V}$ into $\mathbb{P}(\mathcal{V}):=(\mathcal{V}_1,\mathcal{V}_2,\mathcal{V}_3)$. These three subsets represent the different areas labeled in ascending order from the most peripheral to the most central. 
We also consider a penalty for failures, i.e., whenever a user fails to find a shared vehicle within their service area, which represents the quality of the service, and thus the willingness of users to pay for it. Finally, we include a penalty term for cluttering the sidewalks if there are too many vehicles in the same area: this is a widely discussed issue of \acp{MSS}, which may figure in contracts with city governments, as well as increasing fleet management costs.
Hence, the global reward function is composed as:
\begin{equation}
\begin{aligned}\label{eq:global_reward}
R(t) \!=\! 
- \underbrace{\alpha \sum\nolimits_{m=1}^{3} \left[  \phi(m) \sum\nolimits_{i\in\mathcal{V}_m}[a_{i}(t)]_{*} \right]}_{:=\mathrm{reb}_t}
\!-\sum\nolimits_{i\in\mathcal{V}} f_{i}(t) 
\!-\!\xi \sum\nolimits_{i\in\mathcal{V}} \!(|s_{t,i}^v-\mu_{t,i}| - \zeta_{\kappa_i}),
\end{aligned}
\end{equation}
where $\alpha,\xi >0$ are constants, 
$[\cdot]_{*}$ is equal to $0$ if the argument is $0$ and $1$ otherwise, and $\phi:\{1,2,3\}\rightarrow(0,1]$ is a strictly decreasing function. 
Whenever the action $a_{i}(t)$ is nonzero, the product $\tilde{\phi}(m):=\alpha\phi(m)$ is subtracted from the total summation; indeed, the latter quantity can be intended as the cost of carrying out a rebalancing operation for the $m$-th area.
Furthermore, the variable $f_{t,i}$ represents the number of failures over the node $i$ during the considered interval.
Also, the last term of $R(t)$ accounts for the fact that the injection of further vehicles into the network should be penalized proportionally, due to the clutter and fleet maintenance issues discussed above. Such a cost is modeled proportionally to the sum of every mismatch between the current number\footnote{The quantity $s_{t,i}^v$
is upper-bounded by $\sigma_i \gg$ $\mu_{t,i}$, $ \forall t\geq 0$, 
to render the state space finite.}
of vehicles $s_{t,i}^{v}$ $\in [0,\sigma_i]$ and the  expected demand $\mu_{t,i}$ (until the next rebalancing action) at each node $i$, considering some tolerance quantified by a fixed\footnote{Constant  $\zeta_{\kappa_i}$ is a fraction equal to half of the expected arrivals $\mathrm{\bar{a}}_{\kappa_i}$ in all nodes $i$ of the category $\kappa_i$.}
$\zeta_{\kappa_i}\geq 0$, with $\kappa_i \in\{1,2,3\}$ being the index for which $i\in \mathcal{V}_{\kappa_i}$.

\subsection{Reward function augmentation with fairness considerations}
\label{subsec:3b}

Considering a linear combination of profit and a pure fairness metric as an objective function allows us to control the trade-off between economic and fairness concerns.
As the rebalancing cost $\mathrm{reb}_t$
in \eqref{eq:global_reward} is strictly decreasing as we get closer to the city center, while the expected utility of visiting an area is strictly increasing,
a profit-oriented algorithm will tend to rebalance central areas much more often, leading to a lower failure probability. However, spatial fairness consideration lead us to aim at equalizing the failure rates between different areas. 

We can then design a strictly decreasing penalty function $\chi: \{1,2,3\} \rightarrow [-1,1]$ and a fairness weighting parameter $\beta>0$. The product $\tilde{\chi}(m) :=~\beta \chi(m)$ acts as a \textit{temperature}, to measure the degree of importance
that is given to central areas with respect to peripheral ones. The definition of $\chi$ may be arbitrarily chosen by system designers, but its strictly decreasing nature introduces a higher penalty for failures in more peripheral areas, counterbalancing the tendency of profit-maximizing algorithms to privilege central areas.

We can then add the fairness penalty function to the global reward of our \ac{RL} problem:
\begin{equation}\label{eq:our-reward}
    R^{(f)}(t)=R(t)-\beta \sum\nolimits_{m=1}^{3} \left[  \chi(m) \sum\nolimits_{i\in\mathcal{V}_m}f_{i}(t) \right].
\end{equation}

\section{Experiments}\label{sec:exp}

In this section we provide an extensive investigation of different strategies on a case study, to demonstrate the trade-off between performance and equity
and find a viable compromise.

We have considered a medium-sized micromobility sharing system\footnote{The number of nodes per area for each experiment is here reported: $\{60,30,10\}$, 
where we recall that the categories are ordered from remote to central and follow realistic demand patterns~\cite{weinreich2023automatic}. For each of the experiments the training procedure starts with the service areas being subject to the demand parametrized by a Skellam random variable~\cite{skellam1946frequency} with arrival and departure rates $(\lambda_a,\lambda_d)$: $\{(0.3,2),(3.3,1.5),(13.8,7)\}$ in the morning;  $\{(1.5,0.3),(1.5,3.3),(10,13.8)\}$, in the evening.} as an example of dockless \ac{MSS}.
At every hour of the day $t\in\{0,\ldots,23\}$, the number of vehicles present in each area is updated based on the modified MBDP introduced in Section~\ref{sec:sys}.
If at a certain moment a service area is unable to satisfy the demand, i.e. no vehicle is available and there is request for a departure, this is registered as a single failure for that node. The RL agents perform their control actions at 11a.m. and at 11p.m. every day through rebalancing, as described in Section \ref{sec:intro}. The training phase for each strategy is run through $T=10^5$ days and evaluated over $E=10^2$ days.  
The learning rate and epsilon decay are set to $0.01$ and $8.25\cdot 10^{-7}$ respectively; while $\gamma:=0.9$, $\alpha:=20$, $\xi:=0.3$ are chosen. Finally, $\chi$ takes values from the array $y_\chi := [1,.4,-1]$ according to its characterization; $\phi$~takes values from the array $y_\phi :=~[1,.4,.1]$, so that $\phi(m)=~y_\phi[k]$ if $m,k$ are such that $\chi(m)=y_\chi[k]$.

\begin{figure}[t]
  \centering
  \includegraphics[width=8.5cm]{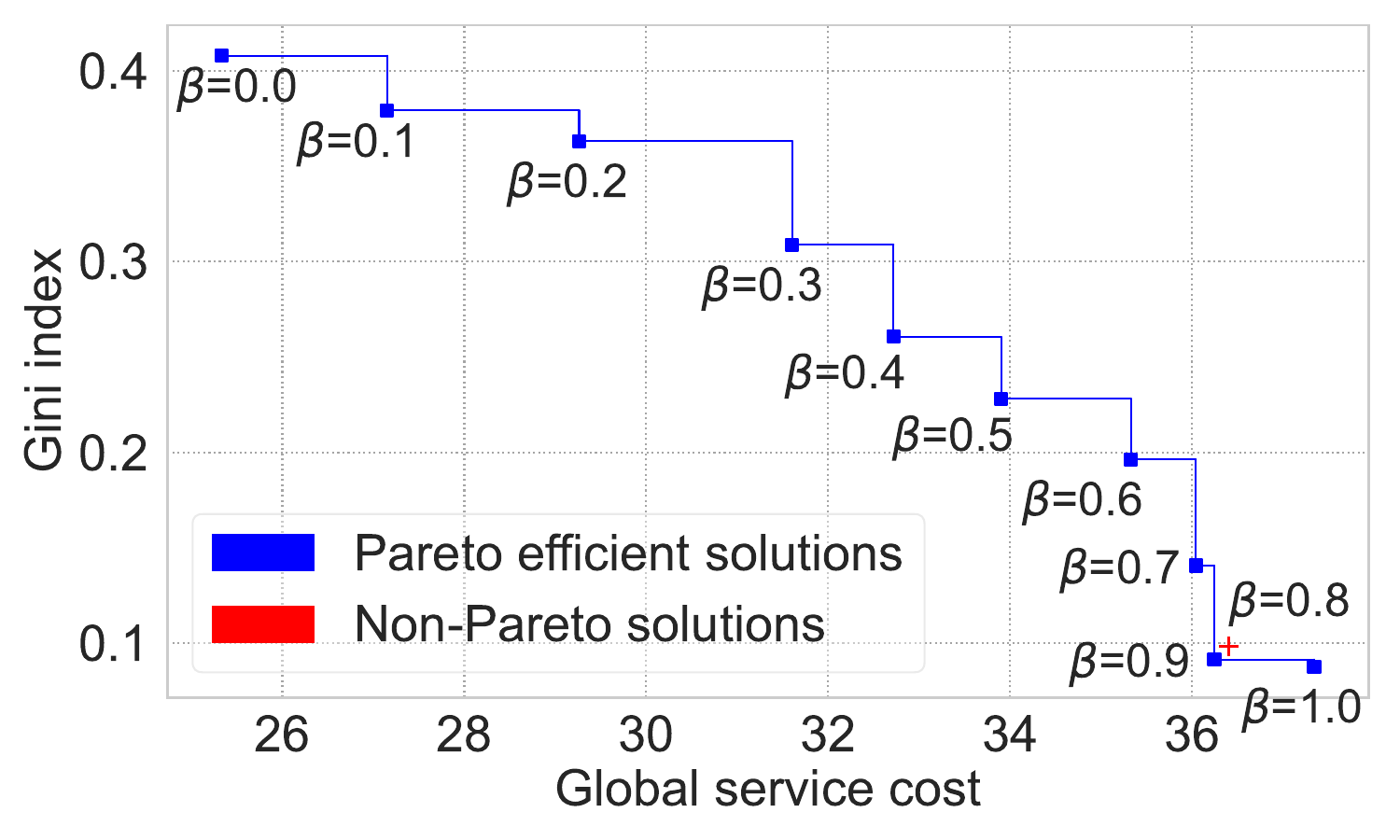}
  \caption{Pareto front for the considered bi-objective optimization problem. The cost minimization and the fairness maximization objectives are represented on the $x$ and $y$ axes, respectively. Each marker corresponds to a different value of $\beta$.}
  \label{fig:pareto}
\end{figure}
\begin{figure*}[t]
    \centering
    \begin{subfigure}{0.25\textwidth}
        \centering
        \includegraphics[width=\linewidth]{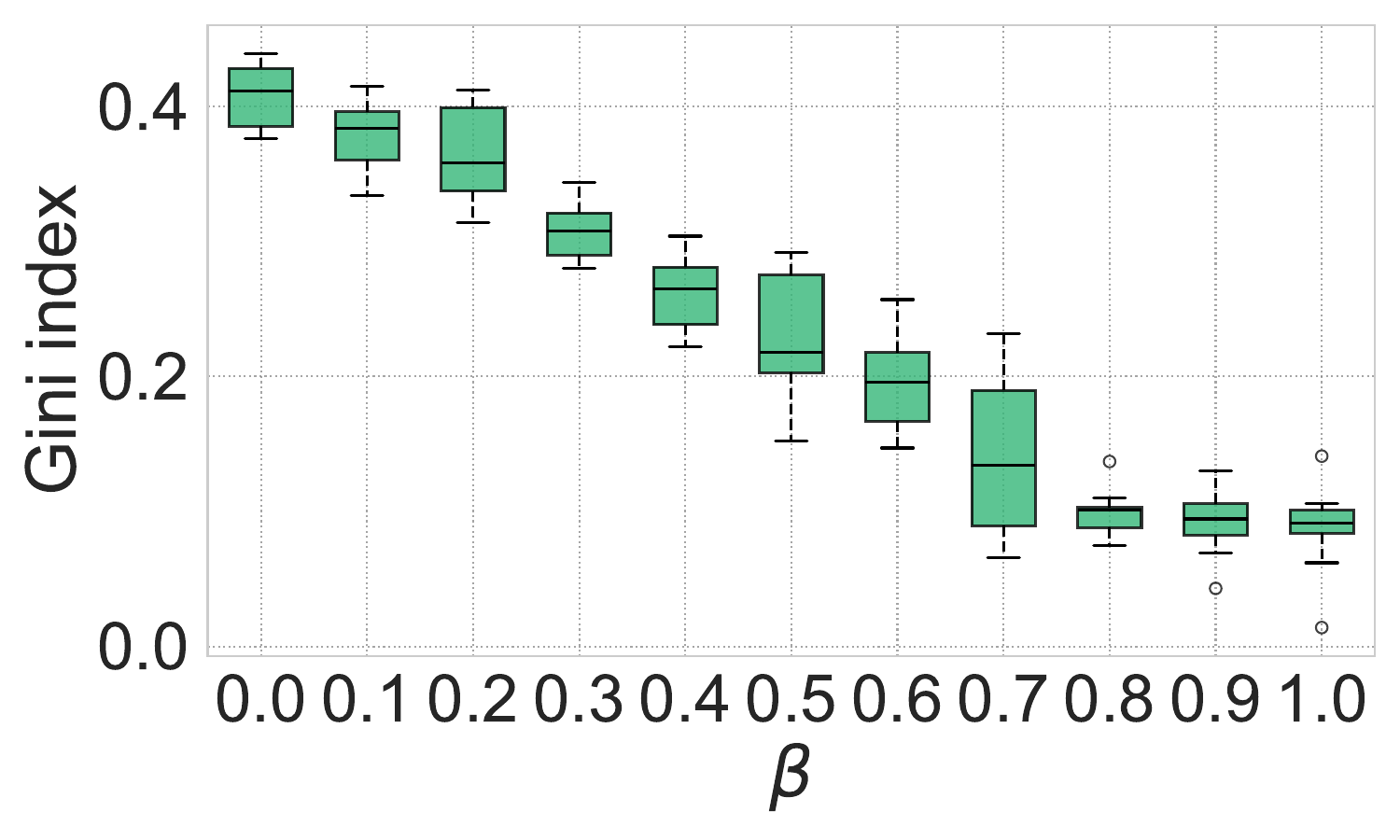}
        \caption{
        $g(x)$ vs. to $\beta$
        }
        \label{fig:boxgini}
    \end{subfigure}%
    \begin{subfigure}{0.25\textwidth}
        \centering
        \includegraphics[width=\linewidth]{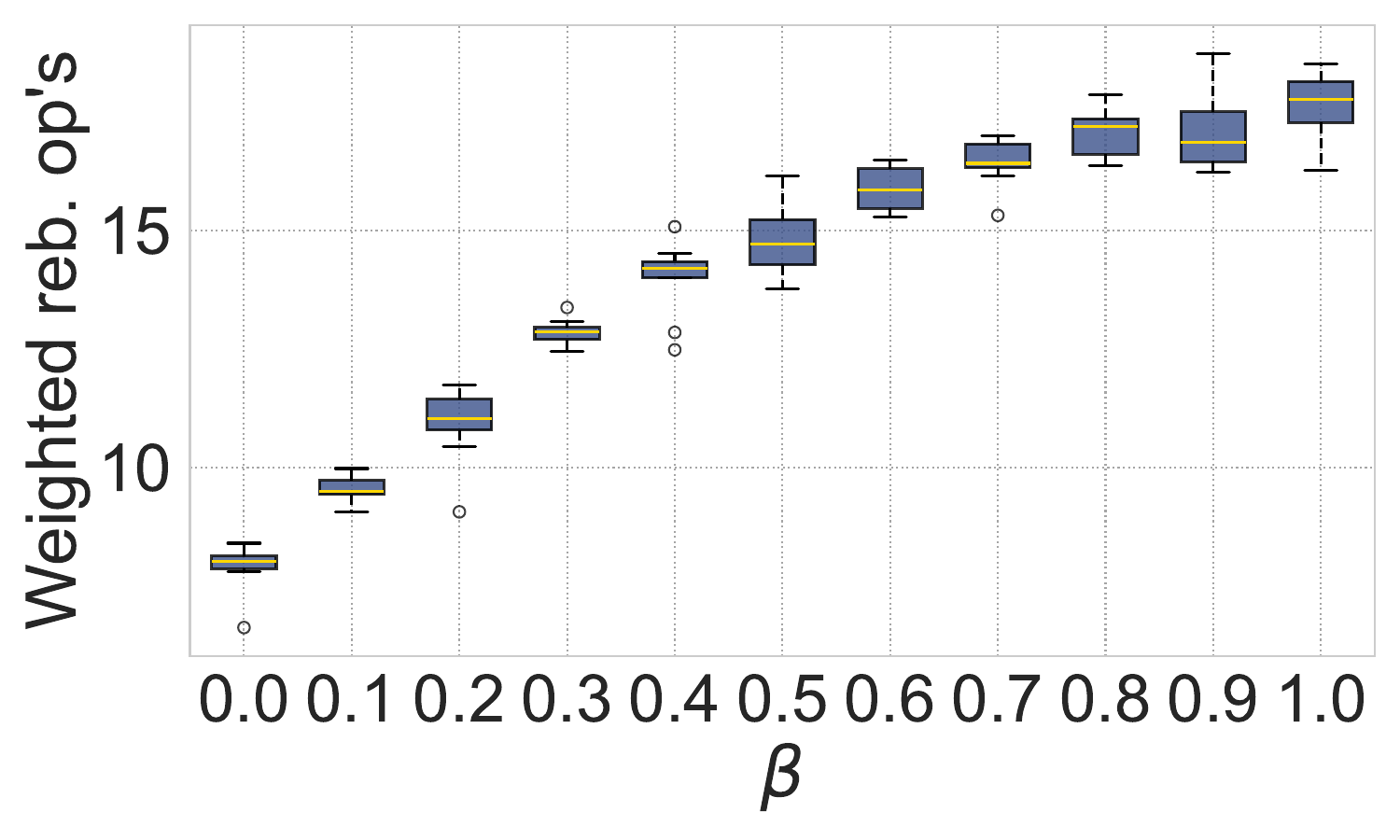}
        \caption{
        $\mathcal{C}_1$ vs. $\beta$
        }
        \label{fig:boxreb}
    \end{subfigure}%
    \begin{subfigure}{0.25\textwidth}
        \centering
        \includegraphics[width=\linewidth]{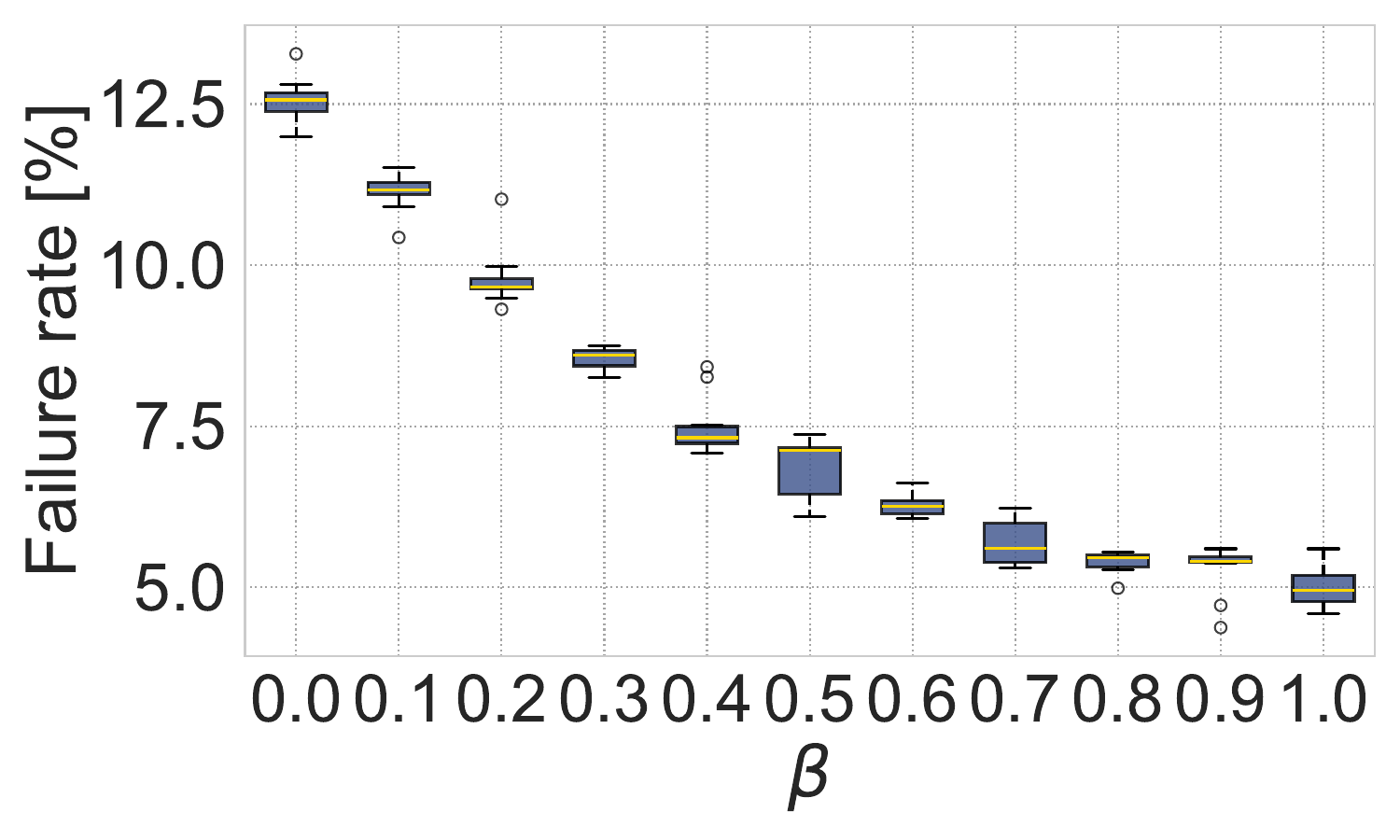}
        \caption{
        $\mathcal{C}_2$ vs. $\beta$
        }
        \label{fig:boxfails}
    \end{subfigure}%
    \begin{subfigure}{0.25\textwidth}
        \centering
        \includegraphics[width=\linewidth]{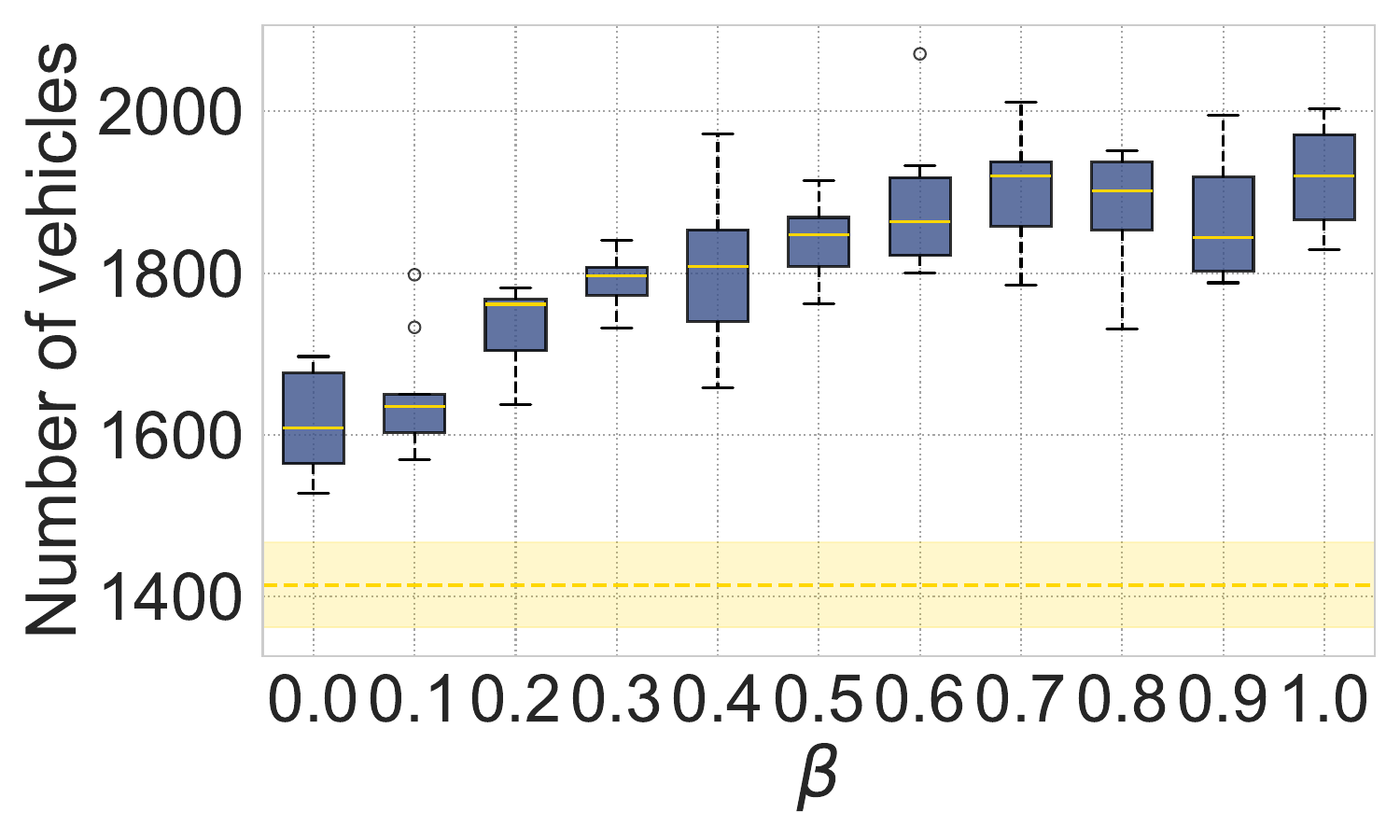}
        \caption{
        $\mathcal{C}_3$ vs. $\beta$
        }
        \label{fig:boxbikes}
    \end{subfigure}
    \caption{Distributions over 10 Monte Carlo runs of fairness and cost-related performance metrics as a function of $\beta$. (d): The yellow dashed line and area indicate respectively average and $97.5$th percentile of the distribution of the initial number of vehicles.}
    \label{fig:boxplots}
\end{figure*}

With the above setup, we analyze the Pareto front of the proposed approach, considering the trade-off between operational costs and fairness.
Upon training and evaluating the algorithm ten times across different seeds for each value of $\beta\in[0,1]$, with step-size $0.1$, the average global service cost and Gini index fairness indicator have been respectively compared on the x and y axes of the Pareto diagram in Figure~\ref{fig:pareto}.
Specifically, we took as global service cost the linear combination of three sources of expenses for the service provider, i.e. $\mathcal{C} :=~\sum_{k=1}^3 \omega_k \mathcal{C}_k$, with $\omega_1:=1$, $ \omega_2:=10$, $\omega_3:=10^{-2}$, where
\begin{equation}\label{eq:C1-C3}
	\mathcal{C}_1 := \text{\small$E^{-1}$} \text{\Large $\Sigma$}_{t=1}^{E} \mathrm{reb}_{t}, \quad
	\mathcal{C}_2 := \text{\small$E^{-1}$} \text{\Large $\Sigma$}_{t=1}^{E} \text{\Large $\Sigma$}_{i\in \mathcal{V}} f_{t,i}/\mu_{t,i}, \quad
	\mathcal{C}_3 := \text{\small$E^{-1}$} \text{\Large $\Sigma$}_{t=1}^{E} \text{\Large $\Sigma$}_{i\in \mathcal{V}} s_{t,i}^v
\end{equation}
respectively denote the number of rebalancing operations, the overall service failure rate and total number of vehicles.

The Pareto-efficient solutions composing the frontier suggest valid choices of implementation, depending both on the desired level of fairness and the costs that the service provider is willing to bear.
Lastly, choosing $\beta = 0.9$ leads to the highest ratio $\rho$ between maximum Gini index decrease ($-77.56\%$) and minimum increase for $\mathcal{C}$ ($+43.05\%$) with respect to applying no equity adjustment, i.e., with $\beta=0$.
To conclude, we examine the distributions of both the fairness indicator and the cost terms~\eqref{eq:C1-C3}
encountered by the service provider as $\beta$ varies. 
It can be appreciated that the monotone trend in the Pareto front of Figure~\ref{fig:pareto} is, as expected, consistent with the decreasing curve of the Gini index depicted in Figure~\ref{fig:boxgini} and the increasing curves of costs $\mathcal{C}_1$ and $\mathcal{C}_3$ respectively shown in Figures~\ref{fig:boxreb},~\ref{fig:boxbikes}. 
On the other hand, as illustrated in Figure~\ref{fig:boxfails}, the decrease of $\mathcal{C}_2$ due to better service in disadvantaged neighborhoods is not enough to compensate for the higher costs needed to perform rebalancing operations (Figure~\ref{fig:boxreb}) and maintain more vehicles in the network (Figure~\ref{fig:boxbikes}).

\section{Conclusions}\label{sec:concl}

This~study~addresses~\ac{MSS}~rebalancing~with a focus on spatial fairness.~A~novel~RL~approach, based on categorizing the network into three city areas, has been designed and tested to evaluate system performance.
Numerical results reveal balanced solutions, characterized~by~a~Pareto front highlighting the trade-off between overall cost and spatial fairness. To our knowledge, this is the first work to explore this trade-off in \ac{MSS} management, rather than just planning, from an RL perspective.
This study is meant to inspire future extensions, such as exploring alternative city partitions and correlations in arrival/departure processes.

\section*{Acknowledgments}

This study was partially carried out within the Italian National Center for Sustainable Mobility (MOST) and received funding from NextGenerationEU (Italian NRRP - CN00000023 - D.D. 1033 17/06/2022 - CUP C93C22002750006).

\bibliographystyle{unsrt}
\bibliography{root.bib}

\begin{thebibliography}{10}

\bibitem{chiariotti2020bike}
F~Chiariotti et~al.
\newblock A bike-sharing optimization framework combining dynamic rebalancing
  and user incentives.
\newblock {\em ACM (TAAS)}, 14(3):1--30, 2020.

\bibitem{smith2015exploring}
C~S Smith et~al.
\newblock Exploring the equity dimensions of {US} bicycle sharing systems.
\newblock Technical report, Western Michigan University. Transp. Res. Center
  for Livable~, 2015.

\bibitem{brown2021docked}
S~Meng et~al.
\newblock Docked vs. dockless equity: Comparing three micromobility service
  geographies.
\newblock {\em Journal of Transport Geography}, 96:103185, 2021.

\bibitem{mooney2019freedom}
S~J Mooney et~al.
\newblock Freedom from the station: Spatial equity in access to dockless bike
  share.
\newblock {\em Journal of Transport Geography}, 74:91--96, 2019.

\bibitem{soja2009city}
E~Soja et~al.
\newblock The city and spatial justice.
\newblock {\em Justice spatiale/Spatial justice}, 1(1):1--5, 2009.

\bibitem{chiariotti2018dynamic}
F~Chiariotti et~al.
\newblock A dynamic approach to rebalancing bike-sharing systems.
\newblock {\em Sensors}, 18(2):512, 2018.

\bibitem{andronov2011markov}
A~M Andronov.
\newblock Markov-modulated birth-death processes.
\newblock {\em Automatic Control and Computer Sciences}, 45:123--132, 2011.

\bibitem{duran2020fair}
D~Duran-Rodas et~al.
\newblock How fair is the allocation of bike-sharing infrastructure? framework
  for a qualitative and quantitative spatial fairness assessment.
\newblock {\em Transportation Research Part A: Policy and Practice},
  140:299--319, 2020.

\bibitem{weinreich2023automatic}
N~A Weinreich et~al.
\newblock Automatic bike sharing system planning from urban environment
  features.
\newblock {\em Transportmetrica B: Transport Dynamics}, 11(1):2226347, 2023.

\bibitem{skellam1946frequency}
J~G Skellam.
\newblock The frequency distribution of the difference between two {Poisson}
  variates belonging to different populations.
\newblock {\em JRSS (A): Statistics in Soc.}, 109(3):296--296, 1946.

\end{thebibliography}

\end{document}